\documentclass[12pt]{article}
\usepackage{amsbsy}
\usepackage{amsmath}
\usepackage{amssymb}
\usepackage{float}
\usepackage{booktabs}
\usepackage{natbib}
\usepackage{caption}
\usepackage{bm}
\usepackage{threeparttable}
\usepackage{fullpage}
\usepackage{mathrsfs}
\usepackage{enumerate}
\usepackage{multirow}
\usepackage{multicol}
\usepackage{array}
\usepackage{color}
\usepackage{epsfig}
\usepackage{graphicx}	
\usepackage{subfigure}
\usepackage{booktabs}
\usepackage{times}
\usepackage[ruled,linesnumbered]{algorithm2e}
\usepackage{booktabs}

\usepackage[titletoc]{appendix}
\usepackage[T1]{fontenc}
\usepackage[utf8]{inputenc}

\usepackage{multirow}

\makeatletter
\def\singlespace{\def\baselinestretch{1}\@normalsize}

\DeclareMathOperator*{\argmax}{arg\,max}
\renewcommand{\hat}{\widehat}

\date{\today}  % Define today as the date
\begin{document}

\title{On Soft Bayesian Additive Regression Trees and
	asynchronous longitudinal regression
	analysis}

\author{Ran Hao,Bai Yang
\\ \small Shanghai University of Finance and Economics,Shanghai\ 200433 \\
 \small E-mail:}
\maketitle

\begin{abstract}
In many longitudinal studies, the covariate and response are often intermittently observed at irregular, mismatched and subject-specific times.
How to deal with such data when covariate and response are observed asynchronously is an often raised problem.
Bayesian Additive Regression Trees(BART)  is a Bayesian non-Parametric approach  which has been shown to be competitive with the best modern predictive methods such as random forest and boosted decision trees.The sum of trees structure combined with a Bayesian inferential framework provide a accurate and robust statistic method.BART variant soft Bayesian Additive Regression Trees(SBART) constructed using randomized decision trees was developed and substantial theoretical and practical benefits were shown.In this paper, we propose a weighted SBART model solution for asynchronous longitudinal data.In comparison to other methods, the current methods are valid under with little assumptions on the covariate process. 
Extensive simulation studies provide numerical support for this solution. And data from an HIV study is used to illustrate our methodology.

\end{abstract}

\noindent {\bf Keywords:}
Soft Bayes Additive Regression Trees,Asynchronous,Longitudinal Data, Non-parametric Regression.

\baselineskip=20pt
%%%%%%%%%%%%%%%%%%%%%%%%%%%%%%%%%%%%%%%%%%%%%%%%%%%%%%%%%%%%%%%%%%%%%%%%%%%%%%%%%%%%%%%%%%%%%%%%%%%%%
\section{Introduction}
\indent
\vspace{-5mm}

Longitudinal data arise in many scientific inquiries, such as epidemiological studies, clinical trials and educational studies. In such studies, data are often collected at subject specific time points and the number of measurements varies across subjects.As an example, in a prospective observational cohort study \citep{2005Cytomegalovirus}, 191 patients were followed for up to 5 years, with HIV viral load and CD4 cell counts measured repeatedly for these patients. Sparse measurements are taken on each variable for each subject and the  viral load and CD4 cell count were obtained at different laboratory on
different days. How to analysis this kind of incomplete longitudinal data is a problem need to be solved.
Some methods are proposed to solve incomplete data problems,such as likelihood based approaches\citep{2000Linear},
inverse probability weighting \citep{1994Estimation},multiple imputation \citep{1996Multiple} and Last observation carried forward.
But these methods need to put stringent assumptions and the inferences are highly dependent on untestable and often implicit assumptions.

In this paper,we focus on the asynchronous longitudinal data defined in \citet{2015Regression} where the measurement times for a longitudinal response and covariate are mismatched.We propose an intuitively appealing weighting non-parametric regression which make full use of the covariate information and change the asynchronous data set into synchronous data.

While regression analysis using estimating equations for  asynchronous longitudinal data has been studied,there has been limited work on the analysis of non-parametric regression.\cite{2010A} proposed a binning method to synchronize the covariate and response.\cite{2016On} proposed a  non-parametric kernel weighting approach using the Last observation carried forward (LOCF) method and show that the estimator is consistent and asymptoticallly but is valid under weak assumptions on the the covariate and observation.

The idea of the kernel weighting is that the closer covariate measured time is from the response measured time,the more relationship between the covariate and response have.So we can weight the pair of covariate and response as an observation by a decreasing function of distance between the two measured time.

Bayesian Additive Regression Trees(BART) is a Bayesian non-parametric approach which has been shown to be competitive with the best modern predictive methods such as random forest and boosted decision trees.The sum of trees structure combined with a Bayesian inferential framework provide a accurate and robust statistic method.BART variant soft Bayesian Additive Regression Trees(SBART) constructed using randomized decision trees was developed and outperform BART model. We choose SBART as the non-parametric estimator for many reasons.First of all,SBART has perfect performance among non-parametric estimator competitors and it is capable of capture the interaction between covariate .Secondly,we analysis the asynchronous longitudinal data based on some assumption that the true function is continuous and the randomized decision tree structure is suitable for this continue situation.And then we can add weight to each cases and apply the algorithm with weight easily with no need to fit estimating equation.

The paper is organized as follows.In section 2,we briefly review the idea of BART and SBART. Section 3 report the main design of weighted  SBART method to analysis asynchronous longitudinal data.Section 4 reports simulation studies that compare the proposed methods  with other methods such as linear interpolation and LOCF.The new method demonstrate improved performance in different settings.An application to an HIV illustrate the practical utility of the methods.Concluding remarks are given in Section 5.

\section{The BART And SBART Algorithm}
We first review those aspects of the BART and SBART methodology  for the understanding of this paper.

\subsection{ The Sum of Trees Model}

For a p-dimensional vector of predictors $X_{i}$ and a response $Y_{i} (1\leq i \leq n)$ ,the BART model posits
\begin{eqnarray}
    Y_{i}=f(X_{i})+\varepsilon_{i},  \varepsilon_{i} \sim N\left(0, \sigma^{2}\right) ,i=1, \cdots, n
\end{eqnarray}
To estimate the non-parametric $f(X)$, a sum of regression trees is specified as

\begin{eqnarray}
    f(X_{i})=\sum_{j=1}^{m} g\left(X_{i} ; T_{j}, M_{j}\right)
\end{eqnarray}
$T_{j}$ is the $j^{th}$ binary tree structure and  $M_{j}=\left\{\mu_{1 j}, \ldots, \mu_{b_{j}}\right\}$is the terminal node parameters associated with $T_{j}$ .$T_{j}$ contains information of which variate to split on ,the splitting value as well as the internal node's location.
$m$ denote the number of trees which is usually set as 200 or 50.

 \subsection{Prior }
 The prior distribution for BART model is $P\left(T_{1}, M_{1}, \ldots, T_{m}, M_{m}, \sigma\right)$.Here we assume that all the trees
 $\left\{\left(T_{1}, M_{1}\right), \ldots,\left(T_{m}, M_{m}\right)\right\}$ are independent with $\sigma$ ,and $\left(T_{1}, M_{1}\right), \ldots,\left(T_{m}, M_{m}\right)$are independent with each other,so we have
  \begin{eqnarray}
  \label{equ:s1}
 \begin{aligned}
P\left(T_{1}, M_{1}, \ldots, T_{m}, M_{m}, \sigma\right) &=P\left(T_{1}, M_{1}, \ldots, T_{m}, M_{m}\right) P(\sigma) \\
& =\left[\prod_{j}^{m} P\left(T_{j}, M_{j}\right)\right] P(\sigma) \\
&=\left[\prod_{j}^{m} P\left(M_{j} \mid T_{j}\right) P\left(T_{j}\right)\right] P(\sigma) \\
&=\left[\prod_{j}^{m}\left\{\prod_{k}^{b_{j}} P\left(\mu_{k j} \mid T_{j}\right)\right\} P\left(T_{j}\right)\right] P(\sigma)
\end{aligned}
  \end{eqnarray}
So we need to specify the prior for $P\left(\mu_{k j} \mid T_{j}\right), P(\sigma),$ and  $P(T_{j})$.
\\
For the convenience of computation, we use the conjugate normal distribution $N\left(\mu_{\mu}, \sigma_{\mu}^{2}\right)$ as  the prior for $\mu_{i j} \mid T_{j}$,$(\mu_{\mu}$,$\sigma_{\mu})$can be derived through computation.
 \\
 The prior for $T_{j}$ is specified by three aspects:
 \begin{itemize}
  \item [1)]
  The probability for a node at depth $d$ to split ,given by $\frac{\alpha}{(1+d)^{\beta}}$.Usually $\alpha$ is set to 0.95 and $\beta$ is set to 2.
  \item [2)]
  The distribution on the splitting variable assignments at each interior node,default as  uniform distribution.
  \item [3)]
  The distribution for cutoff value assignment,default as uniform distribution.
\end{itemize}

We also use a conjugate prior, here the inverse chi-square distribution for $\sigma$,$\sigma^{2} \sim v \lambda / \chi_{v}^{2}$,the two parameters $\lambda$,$v$ can be roughly derived by calculation.
\subsection{Posterior Distribution}

With the settings of priors $(\ref{equ:s1})$,the posterior distribution can be obtained by
\begin{eqnarray}
\label{equ:s2}
\begin{aligned}
P\left[\left(T_{1}, M_{1}\right), \ldots,\left(T_{m}, M_{m}\right), \sigma \mid Y\right] \propto & P\left(Y \mid\left(T_{1}, M_{1}\right), \ldots,\left(T_{m}, M_{m}\right), \sigma\right) \\
& \times P\left(\left(T_{1}, M_{1}\right), \ldots,\left(T_{m}, M_{m}\right), \sigma\right)
\end{aligned}
\end{eqnarray}
The posterior distribution can be obtained by  Gibbs sampling.We can draw m successive 

\begin{eqnarray}
	\label{equ:s3}
	\begin{aligned}
P\left[\left(T_{j}, M_{j}\right) \mid T_{(j)}, M_{(j)}, Y, \sigma\right]
\end{aligned}
\end{eqnarray}
where $T_{(j)}$ and $M_{(j)}$ consist of
all the trees information and parameters except the $j^{th}$ tree.Then draw $P\left[ \sigma \mid \left(T_{1}, M_{1}\right), \ldots,\left(T_{m}, M_{m}\right), Y\right]$ from inverse gamma distribution with explicit expression.
\\

Note that $(T_{j}$, $M_{j})$ depends on  $ T_{(j)}, M_{(j)}$ through
$R_{j}=Y-\sum_{w \neq j} g\left(X, T_{w}, M_{w}\right)$
 ,it's equivalent to draw posterior from a single tree of
\begin{eqnarray}
\label{equ:s4}
P\left[\left(T_{j}, M_{j}\right) \mid R_{j}, \sigma\right].
\end{eqnarray}
We can proceed $(\ref{equ:s4})$ in two steps.First we obtain a draw from $P\left(T_{j} \mid R_{j}, \sigma\right)$,then draw posterior from $P\left( M_{j} \mid T_{j},  R_{j}, \sigma\right)$.In the first step,we have
  \begin{eqnarray} \label{equ:s5}
  P\left(T_{j} \mid R_{j}, \sigma\right) \propto P\left(T_{j}\right) \int P\left(R_{j} \mid M_{j}, T_{j}, \sigma\right) P\left(M_{j} \mid T_{j}, \sigma\right) d M_{j}
  \end{eqnarray}
 ,we call $P\left(R_{j} \mid T_{j}, \sigma\right)= \int P\left(R_{j} \mid M_{j}, T_{j}, \sigma\right) P\left(M_{j} \mid T_{j}, \sigma\right) d M_{j}$ as marginal likelyhood.Because  conjugate normal prior is employed on $ M_{j}$,we can get an explicit expression of the marginal likelihood.

We generate a candidate tree $T_{j}^{*}$ from the previous tree structure $T_{j}$ using  MH algorithm.
we accept the new tree structure with probability

  \begin{eqnarray} \label{equ:s6}
  \alpha\left(T_{j}, T_{j}^{*}\right)=\min \left\{1, \frac{q\left(T_{j}^{*}, T_{j}\right)}{q\left(T_{j}, T_{j}^{*}\right)} \frac{P\left(R_{j} \mid X, T_{j}^{*}\right)}{P\left(R_{j} \mid X, T_{j}\right)} \frac{P\left(T_{j}^{*}\right)}{P\left(T_{j}\right)}\right\}.
  \end{eqnarray}
   $q\left(T_{j}, T_{j}^{*}\right)$ is  the probability for the previous tree $T_{j}$ moves to the new tree $T_{j}^{*}$.
The candidate tree $T_{j}^{*}$ is proposed using four type of moves:
 \begin{itemize}
  \item [1)]
Grow,splitting a current leaf into two new leaves.
  \item [2)]
Prune,collapsing adjacent leaves back into a single leaf.
  \item [3)]
Swap,swapping the decision rules assigned to two connected interior nodes.
  \item [4)]
Change,reassigning a decision rule attached to an interior node.
\end{itemize}

Once we have finished sample new tree structure,we can sample $P\left( M_{j} \mid T_{j},  R_{j}, \sigma\right)$ in explicit normal distribution.

\subsection{ The SBART Algorithm}

A problem shared with tree models is that the resulting estimates of model are step functions,which can introduce error into the model.The BART model archive some degree of smoothing by averaging over the posterior distribution.If the underlying $f_{0}(X)$ is differentiable,we can take advantage of this additional smoothness to get a more accurate model.To introduce smoothness to the model,we can change the decisions made at each node as random rather than deterministic.For example,sample x goes right at branch b of tree $\mathcal{T}$  with probability 
\begin{eqnarray}
\psi(x ; \mathcal{T}, b)=\psi(x ; c_{b}, \tau_{b})=\psi\left(\frac{x_{j}-c_{b}}{\tau_{b}}\right)  
\end{eqnarray}
where
$\tau_{b}$ is the positive bandwidth parameter and $c_{b}$ is the splitting value associated with branch b,$x_{j}$ is the splitting variable.We usually set
\begin{eqnarray}
	\psi(x ; c_{b},\tau_{b})=\left(1+e^{-(x-c_{b}) / \tau_{b}}\right)^{-1}
\end{eqnarray}
so that smaller values of x will have higher probability of going left and vice versa.Note that when $\tau\rightarrow 0$ ,the random decision is equal to the deterministic decision of BART.\citet{linero2018bayesianb} refer to trees constructed using the above random decision rule as soft trees and call this BART variant as SBART.They also showed the substantial theoretical and practical benefits for SBART.

With the definition of the  logistic gating function $\psi(x)$,the probability of going to leaf $\ell$  is
  \begin{eqnarray}
\phi(x ; \mathcal{T}, \ell)=\prod_{b \in A(\ell)} \psi(x ; \mathcal{T}, b)^{1-R_{b}}(1-\psi(x ; \mathcal{T}, b))^{R_{b}}
  \end{eqnarray}
  where $A(\ell)$ is the set of ancestor nodes of leaf $\ell$ and $R_{b}=1$ if the path to $\ell$ goes right at $b$.Here we denote $\phi_{i}$ as the probability vector for the ith sample $x_{i}$ to go to each leaf of the tree.

  We get the marginal likelihood with explicit expression

  \begin{eqnarray}
  	\label{eq:s10}
P\left(R_{j} \mid T_{j}, \sigma,\sigma_{\mu}\right)=\frac{|2 \pi \Omega|^{1 / 2}}{\left(2 \pi \sigma^{2}\right)^{n / 2}\left|2 \pi \sigma_{\mu}^{2} \mathrm{I}\right|^{1 / 2}} \exp \left(-\frac{\|R_{j}\|^{2}}{2 \sigma^{2}}+\frac{1}{2} \widehat{\mu}^{\top} \Omega^{-1} \widehat{\mu}\right),
  \end{eqnarray}
where
  \begin{eqnarray}
\Omega=\left(\frac{\sigma_{\mu}^{2}}{T} \mathrm{I}+\Lambda\right)^{-1}, \quad \Lambda=\sum_{i=1}^{n} \phi_{i} \phi_{i}^{\top} / \sigma^{2}, \quad \widehat{\mu}=\Omega \sum_{i=1}^{n} R_{i} \phi_{i} / \sigma^{2}
  \end{eqnarray}

With the marginal likelihood  we can carry out the MCMC process to derive the posterior distribution sample.

Besides the randomized decision rule mentioned above,the SBART use a sparsity-inducing Dirichlet prior for variables to be selected as splitting rule
  \begin{eqnarray}
s \sim \mathcal{D}\left(a / p, \ldots, a / p\right)
  \end{eqnarray}
so that it can adapted to high dimensional scenario for variable selection.

\section{Analysis Asynchronous Longitudinal Data}
We consider the model 
  \begin{eqnarray}
E\{Y(t) \mid X(t)\}=g\left\{X(t) ,t \right\}
  \end{eqnarray}
where $g$ is an unknown function,$t$ is a univariate time index,$X(t)$ is a time varying covariate,$Y(t)$ is a time varying response.For subject $i=1,\dots,n$,the observation time of covariate process $X_{i}(t)$ and response process $Y_{i}(t)$ are generated from a bivariate counting
process like

  \begin{eqnarray}
N_{i}(t, s)=\sum_{j=1}^{L_{i}} \sum_{k=1}^{M_{i}} I\left(t_{i j} \leq t, s_{i k} \leq s\right)
  \end{eqnarray}
count the number of observation time for response time up to t and covariate time up to $s$ with the response observation times  $\left\{t_{i j}, j=1, \ldots, L_{i}\right\}$  and the covariate observation  $\left\{s_{i k}, k=1, \ldots, M_{i}\right\}$ .when $L_{i}=M_{i}$ and $t_{i j}=s_{i j}, j=1, \ldots, L_{i}$ for each observed response, the problem is the  ordinary synchronous longitudinal problem. \cite{2015Regression} proposed a  kernel weighting method for  generalized linear model with estimating equation for $\beta$ with 
  \begin{eqnarray}
  	\label{eq:s7}
U_{n}^{f}(\beta)=n^{-1} \sum_{i=1}^{n} \int_{0}^{1} \int_{0}^{1} K_{h}(t-s) X_{i}(s)\left[Y_{i}(t)-g\left\{X_{i}(s)^{T} \beta\right\}\right] d N_{i}(t, s)
  \end{eqnarray}
where $K_{h}(t)=K(t / h) / h, K(t)$ is a symmetric kernel function, usually taken to be the Epanechnikov kernel $K(t)=0.75\left(1-t^{2}\right)_{+}$ and $h$ is the bandwidth.The response 
$Y_{i}(t)$ can be a continuous,categorical and the covariate $X_{i}(t)$ can include some time-independent covariate.In our paper we now only consider the situation that all the different covariate are measured at the same time.For other situation,our method can be easily adapted with small modification.But one thing should be noticed that the increasing number of  different covariate time trajectories  may lead to curse of dimension.

In the  asynchronous longitudinal scenario,we are facing with such problem that at time point $s_{ik}$ we have a valid covariate of   $X_{ik}$ without valid information for response.At time point $t_{ij}$,we have a valid response of $Y_{ij}$ without the information for covariate.   

We need the following conditions.
\begin{enumerate}[1)]
	\item $N_{i}(t, s)$ is independent of $(Y_{i}, X_{i})$.
	\item The true function $g(X(t),t)$ is Lipschitz continue for $X(t)$ with constant $k_{1}$.
	\item The $X(t)$ process is Lipschitz continue for t with constant $k_{2}$.
\end{enumerate}

The predict bias is $|g(X(t_{ij}),t_{ij})-g(X(s_{ik}),t_{ij})| \leq k_{1} |X(t_{ij})-X(s_{ik})| \leq k_{1} k_{2}|t_{ij}-s_{ik}|  $,so with a weight function of $\left(1-(\frac{t_{ij}-s_{ik}}{h})^{2}\right)_{+}$
we can reduce the influence of the predict bias and get a acceptable estimation.

We adapt the estimation problem $(\ref{eq:s7} )$ to a optimization problem $L_{i}$ observations of

\begin{equation}\label{eq:s8}
	\begin{split}
		 \hat{g}&= \mathop{\argmax}_{g}\sum_{i=1}^{n} \int_{0}^{1} \int_{0}^{1} K_{h}(t-s) \left[Y_{i}(t)-g(X_{i}(s),t)\right] ^{2} d N_{i}(t, s) \\
		& = \mathop{\argmax}_{g} \sum_{i=1}^{n} \sum_{j=1}^{L_{i}} \sum_{k=1}^{M_{i}}  K_{h}(t_{ij}-s_{ik}) \left[Y(t_{ij})-g(X_{i}(s_{ik}),t_{ij})\right] ^{2} \\
		& = \mathop{\argmax}_{g} \sum_{i=1}^{n} \sum_{j=1}^{L_{i}} \sum_{k=1}^{M_{i}}  K_{h}(t_{ij}-s_{ik}) \left[Y_{ij}-g(X_{ik},t_{ij})\right] ^{2} \\
	\end{split}
\end{equation}
We then turn the optimization problem of $(\ref{eq:s8} )$ into a weighted SBART model.So we actually draw posterior sample instead of solve the optimization problem.$(\ref{eq:s8} )$ can be interpreted this way.For subject i,we have  $L_{i}$ observations of response and $M_{i}$ observations of covariate.Then we have $L_{i}*M_{i}$ combinations of response  $Y_{ij}$ and covariate $X_{ik}$.We can assume that with $X_{ik}$ and $t_{ik}$
we can get a response of $Y_{ij}$ with confidence $K_{h}(t_{ij}-s_{ik})$ ,that is the more close 
$t_{ij}$ and $s_{ik}$ is ,the more confidence we have to get the response $Y_{ij}$.Here we set  
$K_{h}(t)=K(t / h) =\left(1-\frac{t}{h}^{2}\right)_{+}$.So for the synchronous situation,we can set the bandwidth parameter small enough so that only the synchronous pair of response and covariate are include in the bandwidth range.Using $(\ref{eq:s8} )$ we can get the same result with the synchronous data set. 

In $(\ref{eq:s8} )$ we ignore the information of $s_{ik}$ and only use it in the weight $K_{h}(t_{ij}-s_{ik})$.Another option is that we get the estimation by
\begin{equation}\label{eq:s9}
	\begin{split}
		\hat{g}	& = \mathop{\argmax}_{g} \sum_{i=1}^{n} \sum_{j=1}^{L_{i}} \sum_{k=1}^{M_{i}}  K_{h}(t_{ij}-s_{ik}) \left[Y(t_{ij})-g(X_{i}(s_{ik}),t_{ij})\right] ^{2} \\
		& = \mathop{\argmax}_{g} \sum_{i=1}^{n} \sum_{j=1}^{L_{i}} \sum_{k=1}^{M_{i}}  K_{h}(t_{ij}-s_{ik}) \left[Y_{ij}-g(X_{ik},t_{ij},s_{ik})\right] ^{2} \\
	\end{split}
\end{equation}
The only difference  between two algorithms is that in the later algorithm we use two time trajectories in the estimation process.It can outperform the first algorithm in some situation.We denote the first algorithm as single trajectory (ST) and the later  algorithm as double trajectory (DT).

Now we focus on how to get the non-parametric  estimator  	$\hat{g}$ .Considering the good performance of SBART,we use SBART  to fit the nonparametric model and  use $K_{h}(t_{ij}-s_{ik})$ as the  weighting of this virtual pair of response and covariate $(Y_{ij},X_{ik},t_{ik},s_{ik})$.In the previous section ,we have the marginal likelihood of $(\ref{eq:s10} )$.After we add weight $W_{i}$ to the $i^{th}$case,we have 

  \begin{eqnarray}
	\label{eq:s11}
	P\left(R_{j} \mid T_{j}, \sigma,\sigma_{\mu},W_{j}\right)=\frac{|2 \pi \Omega|^{1 / 2}}{\left(2 \pi \sigma^{2}\right)^{n / 2}\left|2 \pi \sigma_{\mu}^{2} \mathrm{I}\right|^{1 / 2}} \exp \left(-\frac{\|R_{j}\|^{2}}{2 \sigma^{2}}+\frac{1}{2} \widehat{\mu}^{\top} \Omega^{-1} \widehat{\mu}\right),
\end{eqnarray}
where
\begin{eqnarray}
	\Omega=\left(\frac{\sigma_{\mu}^{2}}{T} \mathrm{I}+\Lambda\right)^{-1}, \quad \Lambda=\sum_{i=1}^{n} W_{i}\phi_{i} \phi_{i}^{\top} / \sigma^{2}, \quad \widehat{\mu}=\Omega \sum_{i=1}^{n} W_{i} R_{i} \phi_{i} / \sigma^{2}
\end{eqnarray}
Besides the modification of marginal likelihood,we also have to change the sample process of $\sigma$.
We run the MCMC steps with this version of weighted marginal likelihood and can finally derive the result for the weighted Soft BART.

Compared to commonly used method such as  Last observation carried forward (LOCF) ,our weighted method
can  bring more information into consideration and have more sample size to build the model.For example for subject j,if we apply LOCF method,we may get less than  $L_{i}$ pairs of response and covariate into consideration.Within these pairs of response and covariate there may exist some cases that the response and covariate is far away from each other for which actually we have little confidence.In our weighted method,when the bandwidth $h$ is wider enough,we even can get $L_{i}*M_{i}$ pairs of response and covariate with the confidence consideration of the time distance between response and covariate.With proper bandwidth we can ignore the cases that the response and covariate time is too far for us to have confidence with it.

Then we turn our attention to the bandwidth $h$.When the $h$ is small enough we are ignoring the data which is asynchronous.Now the only problem is that we may not get enough data to accurately estimate the model.As mentioned before we actually estimate $Y_{ij}$ with $g(X_{ik},t_{ij})$,the gap between $g(X_{ik},t_{ij})$ and  $g(X_{ij},t_{ij})$ is the bias introduced into the estimation process.To find the best suitable bandwidth is how to balance the bias and estimation error.The bandwidth depends on many condition such as the total sample size,the distance distribution and the  fluctuation of the covariate and its influence to the response.

One way to set default bandwidth is to make the sample size comparable to other competitor.For example,we will include about $\sum_{i=1}^{n} L_{i} $ cases in LOCF and with our weighted algorithm we have $\sum_{i=1}^{n} L_{i} * M_{i} $ cases ,so we sort the cases by distance and set the bandwidth so that the closest $\sum_{i=1}^{n} L_{i} $ cases are included in the model.By this means we can keep the sample size comparable to other methods.

Best bandwidth is searched with grid. How to judge which bandwidth is the best.Under the weighted setting,we can't directly judge with the  estimated $\hat{\sigma^{2}}$.Linear   interpolation and cross validation are combined together to find the optimal bandwidth.First we use linear interpolation to predict the missing $X_{ij}$ and calculating the distance to the nearest $X_{ik}$.Then we  select out the nearest $D$ cases and use $(X_{ij},t_{ij})$ as part of the test data.For $D$ we always choose a small number to make sure these sample size is enough for test the effect and the distance is not so far that the bias can be omitted.We use cross validation by subjects to get the estimation for $(X_{ij},t_{ij})$ as $\hat{Y_{ij}}$.So we can use the statistics  $\sum_{ij=1}^{D} (\hat{Y_{ij}} -Y_{ij} ) ^{2} / D $ to judge the optimal bandwidth.If the sample size is sufficient,we can even divide the $D$ cases into smaller groups and use the group result will make a more robust judgment.
\section{Simulation Studies And A Real Example}
\subsection{Simulation Studies}
\subsubsection{Comparisons With Competing Methods}

We conducted extensive simulation studies to evaluate the performance of the weighted SBART algorithm in contrast with other competitors.We generate 100 data set,each consist of n=300 or 900 subjects.We split $70\%$ of the subjects into training data set and left the $30\%$ of the subjects as the   The numbers of response $Y(t)$ minus 1 and the numbers of covariate $X(t)$ minus 1 are generated from Poisson distribution with intensity rate $5$ to make sure there will not be a subject with no response or covariate. The observation times for the response
and covariate are generated from uniform distribution U(0, 1) independently.
The covariate process generated from Gaussian process, with values at observed time points being multivariate normal with zero mean  , variance 1 and correlation $e^{-\left|t_{i j}-t_{i k}\right|}$, where $t_{i j}$ and $t_{i k}$ are $j$ th and $k$ th measurement time for the response, both on subject
i. The response process was generated from
$$
Y(X,t)=\alpha(t)+X(t) \beta+\epsilon(t)=f(X,t)+\epsilon(t)
$$
where $\alpha_{t}$ is the non-parametric part, $\beta_{1}$ is regression coefficient and $\epsilon(t)$ from  Gaussian process, with zero mean, variance 1 and $\operatorname{cov}\{\epsilon(s), \epsilon(t)\}=2^{-|t-s|}$. Once the response is generated,we remove the corresponding measurements to make the data asynchronous in the training data.We keep a synchronous data just to check the effect the model.We set four response process by

\begin{eqnarray}
	\begin{aligned}
	\label{eqn.s13}
	f_{1}(X,t) :  \alpha(t)&=\sin (2 \pi t)   & \beta=-2 \\
	f_{2}(X,t) :  \alpha(t)&=\sqrt{t}   & \beta=-2 \\
	f_{3}(X,t) :  \alpha(t)&=0.4 t+0.5   & \beta=-2 \\
	f_{4}(X,t) :  \alpha(t)&=\sin (2 \pi t)   & \beta=-20 
	\end{aligned}
\end{eqnarray}	

To gauge how well our weighted SBART(WSB) perform on the sample data set,we compare its performance in test data set with LOCF-B,LOCF-S,LI,NWT,WTSTD.LOCF-B denotes the method  we adapt the LOCF algoritm to synchronize data and use BART model to fit the unknown function.This competitor is introduced to make comparation with LOCF-S in which the SBART is adapted along with the LOCF.LI denotes  linear interpolation in which we use  linear interpolation to fill the missing covariate information.NWT denotes the competitor in which we only control the bandwidth to default setting and delete cases that the time distance is beyond the bandwidth.In the NWT situation,we don't weight the cases.In the BART package,we can specify each case's weight for heterogeneous BART which we denote WTSTD.As mentioned before,we have double trajectory weighted Soft BART(WSB-DT) and single trajectory weighted Soft BART(WSB-ST).Because we know the value of true function $f(x,t)$,we assess the fit using data in the test set 
$\mathrm{RMSE}=\sqrt{\frac{1}{N} \sum_{i=1}^{N}\left(\hat{f}\left(x_{i},t_{i}\right)-f\left(x_{i},t_{i}\right)\right)^{2}}$ where $N$ is the sample size of the test data set.For each method we thus obtained 100 such RMSE values.

\begin{figure}[htb]
	\begin{center}
		\includegraphics[scale=0.4]{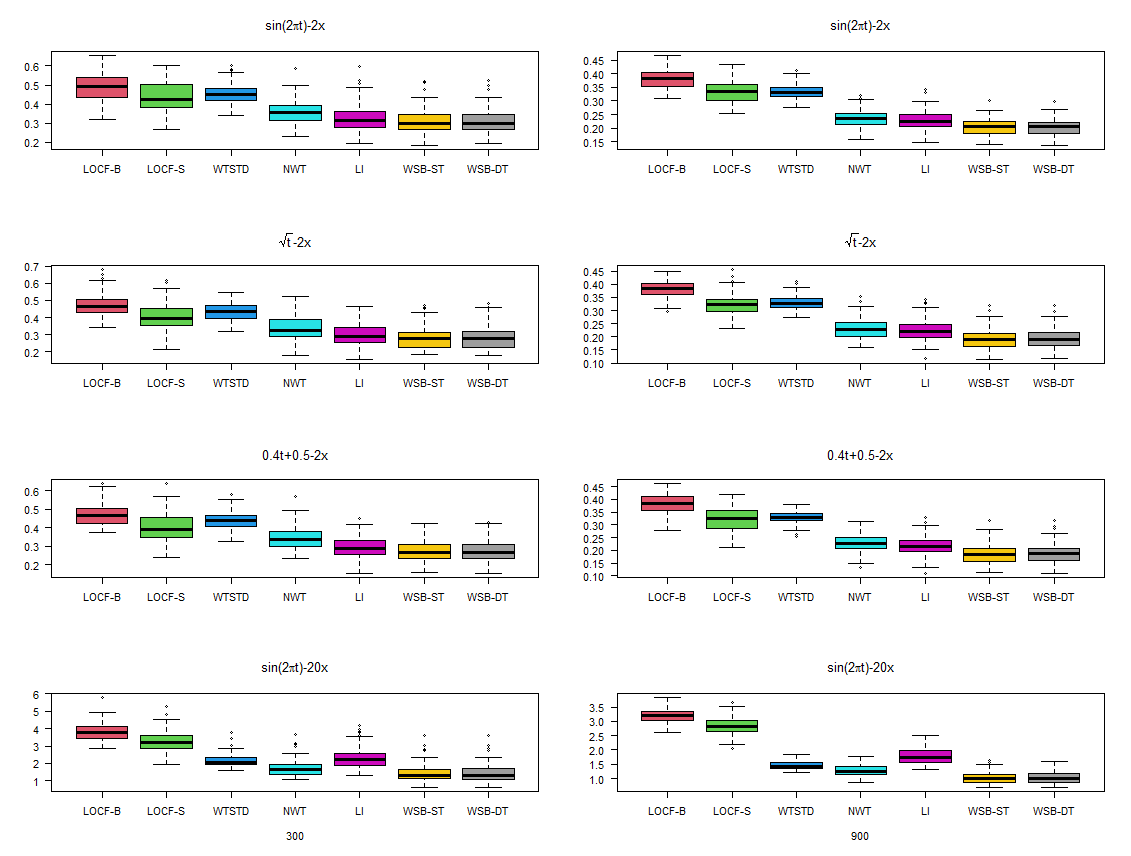}
	\end{center}
	\caption  {  \small{\it{ Predictive comparisons for Weighted Soft BART and its competitors.Each box plot represents 100 RMSE results. }}}
	\label{FIG1}
\end{figure}

Figure 1 displays box plots of the 100 RMSE values for each method for different settings.We can see that the SBART consistently outperform BART with LOCF algorithm.It just heighten our confidence in SBART so all the other competitors except WTSTD are based on SBART .LOCF is actually a bad choice to synchronize data.That can be understand as that if the covariate have some tendency,carrying the last valid covariate forward will introduce some obvious bias into the estimation.Weight cases for the heterogeneous BART seems can't improve the performance.It seems can reduce the variation but that is not our main purpose.We use default bandwidth to filter out all less confident cases.It outperform LOCF-S as we expect.When the subject sample size is small as 300 for the first 3 functions LI is  better than NWT but when subject sample size increase to 900 NWT and LI seems to have equal performance.For function 	$f_{4}(X,t)$ ,we increase the coefficient of X from 2 to 20 so if we wrongly predict the $X_{ij}$ by 0.5,the bias will increase to 10.Under this function setting,the performance of LI is even worse than WTSTD.WSB-ST and WSB-DT seems have similar performance at all settings and outperform all competitor.

\subsubsection{Bandwidth Search}
\begin{figure}[htb]
	
	\begin{center}
		\includegraphics[scale=0.6]{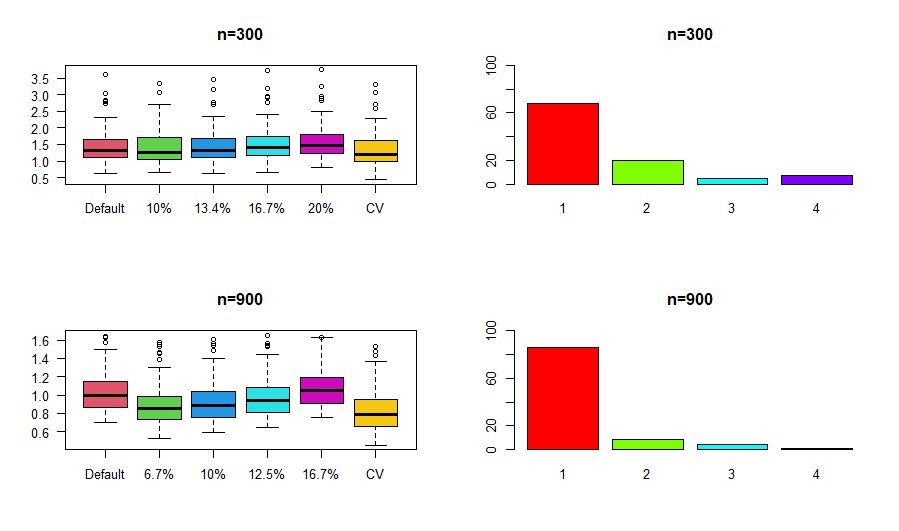}
	\end{center}
	\caption  {  \small{\it{ Bandwidth searching process }}}
	\label{FIG2}
\end{figure}
We show the  our cross validation process for bandwidth $h$ for function 4.We denote the percentage of valid cases included in the model as bandwidth measurement.We start from the default bandwidth of $16.7\%$.We use 10-fold cross validation in the linear interpolation part to find the optimal bandwidth and  RMSE of the test data is used to check the validity of our bandwidth search process.

When subject sample size is 300,we keep the linear interpolation predict sample which the time distance is within the nearest  $8.4\%$ of the linear interpolation samples and set search grid  as $(10.0\%,13.4\%,16.7\%,20.0\%)$.When subject sample size is 900,we keep the linear interpolation predict sample which the time distance is within the nearest  $5\%$ of the linear interpolation samples.Because we assume when we have enough sample to include in the model the bandwidth may tend to be small,so the search grid   $(6.7\%,10\%,12.5\%,16.7\%)$.For robust purpose,we divide the linear interpolation data set into 5 groups and give the smallest distance group more voting score and select the bandwidth with the largest voting score.

Figure $\ref{FIG2}$ displays box plots of RSME result for the search grid and the final decided bandwidth.When subject sample size is 300,at the top of figure $\ref{FIG2}$,we can see that from $20.0\%$ down to $10.0\%$ the RMSE slightly decrease which means $10.0\%$ should be our optimal bandwidth.But the difference of the RMSE at different grid points is not so big ,so from the right part we can see about more than $30\%$ chose the non-optimal bandwidth.Through bandwidth search we can reduce the mean RMSE from 1.448 to 1.345.

When subject sample size is 900,at the bottom of figure 2,we can see the difference between the RMSE of the grid point is more obvious and the optimal bandwidth should be  $6.7\%$.It confirm our assumption that the more samples we get,the narrower bandwidth is preferred.From the right part we can see more than $80\%$ chose the optimal bandwidth.Through bandwidth search we can reduce the mean RMSE from 1.03 to 0.83 which is a remarkable improvement for the estimation.

\subsubsection{Single Trajectory And  Double Trajectory }
\begin{figure}[htb]
	
	\begin{center}
		\includegraphics[scale=0.4]{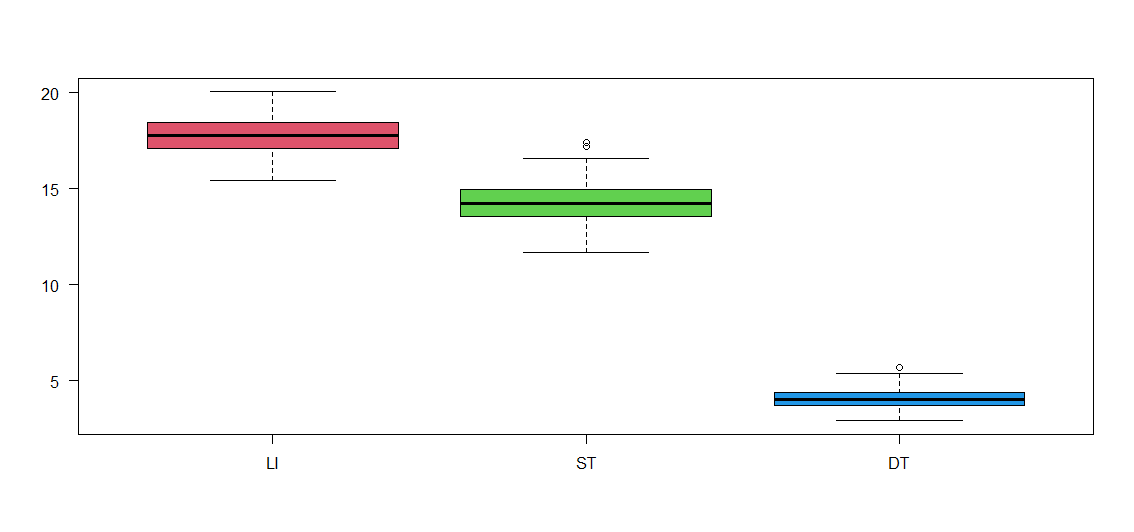}
	\end{center}
	\caption  {  \small{\it{ Single Trajectory .vs.  Double Trajectory }}}
	\label{FIG3}
\end{figure}
In the previous part,we can see that single trajectory(ST) and double trajectory(DT) have almost the same performance.In this section,we show the double trajectory will act better for some situation.When the covariate is just Gaussian process and has no correlation with the time variable.whether we choose ST or DT won't make any difference.But for the situation that covariate is correlate with time variable,we can derive better performance when we include the covariate time information.We keep all the setting for function 4 unchanged except that 
$X=GP(t)+5sin(10\pi t)$ which GP(t) stands for a Gaussian process  with zero mean  , variance 1 and correlation $e^{-\left|t_{i j}-t_{i k}\right|}$.

Figure $\ref{FIG3}$ displays box plot of RMSE result for the linear interpolation ,DT and ST.For this setting of function,due to the high frequency sine part of the covariate process,it's dangerous to use linear interpolation.So the performance of linear interpolation is not good.Because of  the volatile  fluctuation of the covariate,ST algorithm works bad too.Here the ST bandwidth is derived through cross validation.The search process tent to find a small bandwidth.But for the Double Trajectory,we can derive information from the covariate time variable that make a better estimation than ST.The bandwidth derived by DT tent to a large bandwidth so that it can get more information by include more samples. 

\subsubsection{Lagged analysis}
\begin{figure}[htb]
	\begin{center}
		\includegraphics[scale=0.4]{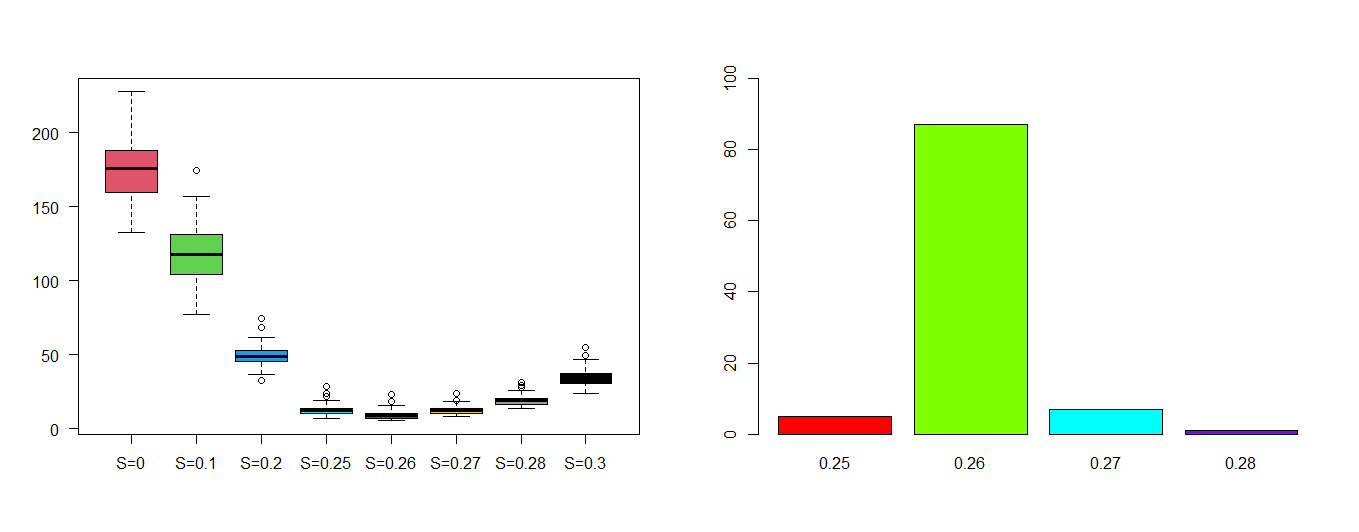}
	\end{center}
	\caption  {  \small{\it{ Lag Dectection Process }}}
	\label{FIG4}
\end{figure}
For the asynchronous or synchronous data ,if lag effect of the covariate exist,we denote the lag time as $t_{lag}$, 
We consider the model 
\begin{eqnarray}
	E\{Y(t) \mid X(t-t_{lag}),t\}=g\left\{X(t- t_{lag}),t\right\}
\end{eqnarray}
For example ,the period of time between the time when patients are infected by Covid-19 and the time when they show some symptoms can be treated as lag time.When lag effect exists and it actually turn a synchronous problem into a asynchronous problem.
For this situation we can use the same search method as to find the optimal bandwidth to find the lag time $t_{lag}$.Here we still use the linear interpolation predict error as the statistics to find the best $t_{lag}$. 
Function 5 is defined as 
$$
f_{5}(X,t)=10*sin(2\pi t)-20x(t-t_{lag})
$$
We set $t_{lag}=0.26$.A grid search $(0,0.1,0.2,0.25,0.26,0.27,0.28,0.3)$ for the lag time $t_{lag}$ is adapted.

Figure $\ref{FIG4}$ displays box plot of RMSE result for the lag search grid and the final decision made for 100 data sets.We can see that we can easily identify the lag effect around 0.26,but due to asynchronous data structure,we can't avoid making small mistaken .After all,we can detect the true lag parameter $0.26$ for more than $80\%$ of the data sets.  
\subsection{Application to HIV data}

A total of 190 HIV patients were followed from July 1997 to September 2002. 
Details of the study design, methods and medical implications are given in\citep{2005Cytomegalovirus}. 
\begin{figure}[htb]
	\begin{center}
		\includegraphics[scale=0.4]{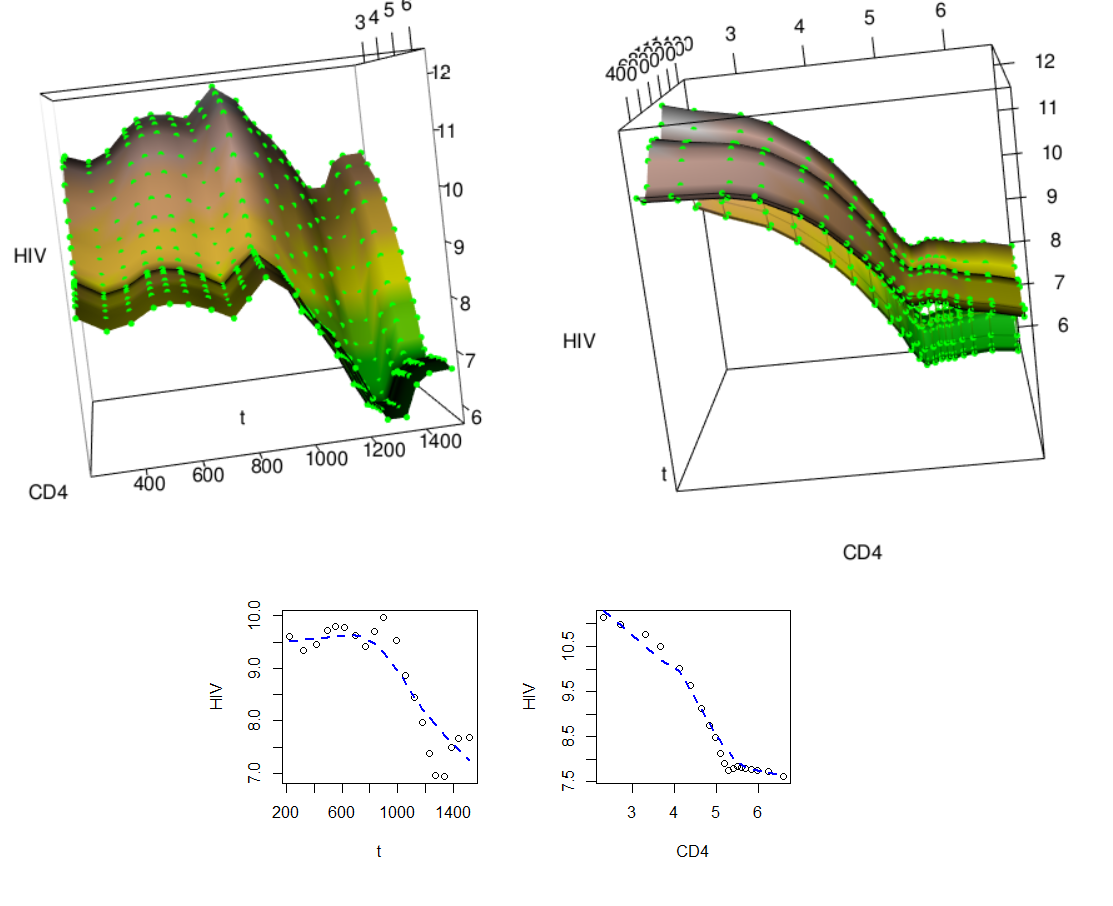}
	\end{center}
	\caption  {  \small{\it{ The non-parametric model for HIV data and its marginal effect on t and CD4 count.Default Bandwidth is $149$. }}}
	\label{FIG5}
\end{figure}
In this study, all patients were scheduled to have their measurements taken during semiannual visits, with HIV viral load and CD4 cell counts obtained separately at different laboratories.
Because many patients missed visits and the HIV infection occurred randomly during the study,
there are unequal numbers of repeated measurements on viral load and CD4 cell count and
there are different measurement times for the two variables. 
These data are sparse and asynchronous.

In our analysis, we took the CD4 cell counts as the covariate and HIV viral load as the response. 
Both CD4 cell count and HIV viral load are continuous variables with skewed distribution, we log-transformed these variables before the analysis.

First we clean the data and delete invalid cases.After data cleaning,we have 181 valid subjects with 563 valid response observations and 723 valid covariate  observations.Combine all the response and covariate of the same subject together we have 3128 cases.The bandwidth is selected with $20\%$ of the  cases are included in the model,now the bandwidth is $149$.

Top half of figure  $\ref{FIG5}$ displays the non-parametric model obtained with our weighted SBART algorithm.At the bottom is the marginal effect for time variable $t$ and covariate CD4.From the marginal effect of CD4,obvious negative relationship between CD4 counts and HIV viral load can be observed which is mentioned in the  work of  \citep{2015Regression}.We can find that the linear model is not suitable especially when CD4 is above 5.And the marginal effect of time variable,we can see a drop of the HIV virus loading after 1000 days which is also presented in \citep{2017Analysis}. The fluctuation of the marginal estimation around 1400 days can be observed and why this happens is an interesting problem.Compare the marginal effect with non parametric model we can see that the pattern can be extended along the coordinate for both variables.The interaction effect of the two variables is not so obvious just by observation.  

\begin{figure}[htb]
	
	\begin{center}
		\includegraphics[scale=0.6]{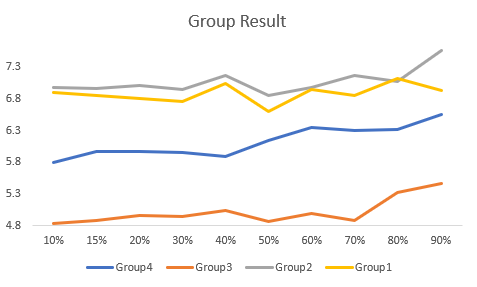}
	\end{center}
	\caption  {  \small{\it{ Bandwidth selection process}}}
	\label{FIG6}
\end{figure}

For the previous part,default bandwidth is adapted.We then try a bandwidth searching process in which cases with time distance between the covariate and response below 75 days are used to predict its covariate with linear interpolation and keep these samples to test performance of bandwidth.We divide the test cases into 4 groups so that we try to get a more robust choice of bandwidth. 

\begin{figure}[H]
	\begin{center}
		\includegraphics[scale=0.5]{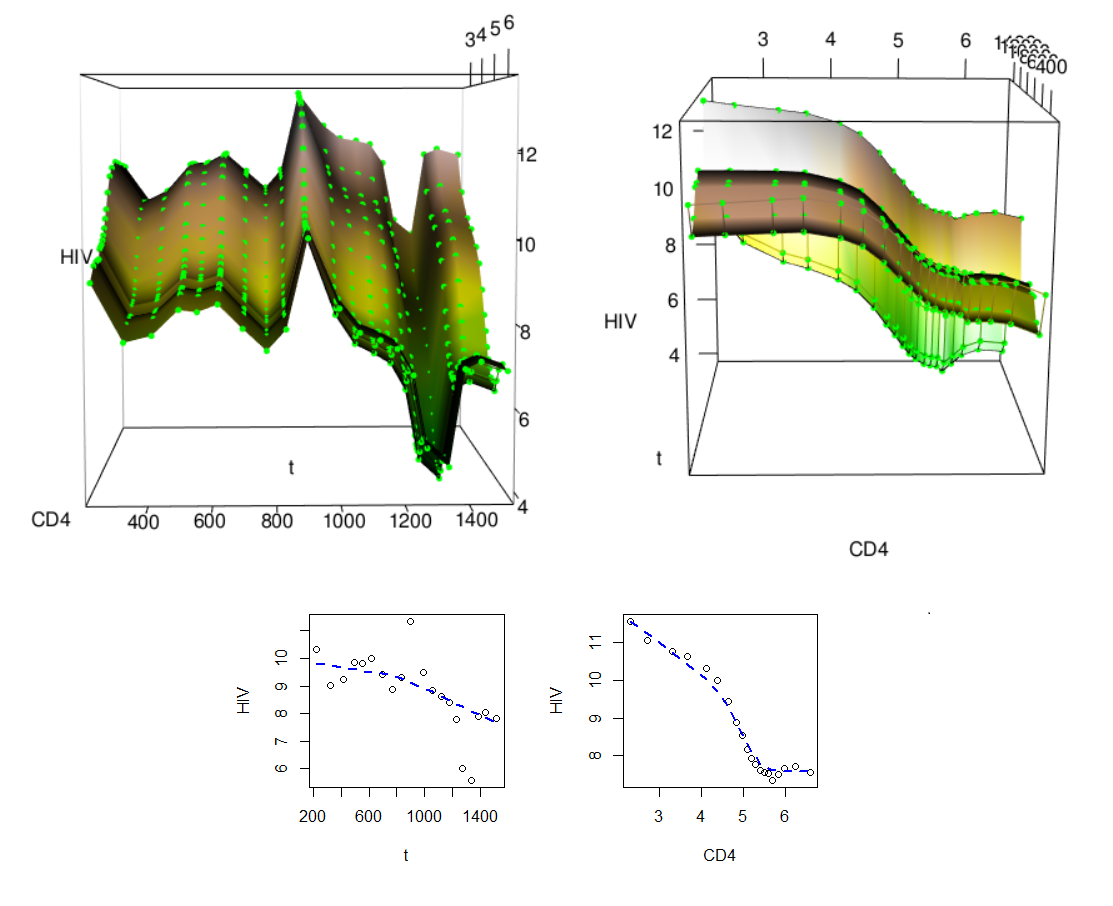}
	\end{center}
	\caption  {  \small{\it{Non-parametric model for HIV data and its marginal effect on t and CD4 with bandwidth of 324. }}}
	\label{FIG7}
\end{figure}

From Figure $\ref{FIG6}$,we can see that at $50\%$ we obtain the lowest test error for 3 groups out of 4 groups.$50\%$ means a bandwidth is selected so  $50\%$ of the samples are included in the model.So we choose the corresponding bandwidth $324$.With this bandwidth,we get a similar result compared to the default bandwidth which can be seen in Figure $\ref{FIG7}$.
With a wider bandwidth,the non-parametric model seems more steep and fluctuation of the marginal effect of time variable around days 900 and 1400 should be inspected carefully.

\section{Concluding remarks}
In this paper ,we propose a non-parametric model to fit asynchronous longitudinal data.Instead of estimating equation,we try weighted SBART algorithm.
This allows us to explore the curvature and nonlinear effect of longitudinal response.Along with this algorithm,we design a linear interpolation cross validation method to find the optimal bandwidth and show good performance in data experiments.Numerical studies support our algorithm and show its power in contrast to other competitors.Our double trajectory method outperform single trajectory method  when covariate and time is correlated and we can obtain information from the covariate to get a better estimation of the response.This algorithm is capable of solving lag issues in which case the synchronous longitudinal data is asynchronous longitudinal data and our method can identify the lag time with the same process we choose bandwidth.New algorithm is adapted to the HIV data and find fitting the covariate part with linear model is not a good choice especially when CD4 count is above 5.5.  

Our method is based on a working independence assumption which means we don't make use the correlation structure of the longitudinal data.The bias introduced from asynchronous background can not easily be identified,so we have to treat it as part of error.With the asynchronous bias ,the original correlated error structure is hard to discriminate.How to make use of the correlated structure and derive a better estimation  is left for future discussion.

\newpage

\vskip20pt
\def\refhg{\hangindent=20pt\hangafter=1}
\def\refmark{\par\vskip 1mm\noindent\refhg}
\bibliographystyle{plainnat}

\bibliography{ASYN}

\end{document}